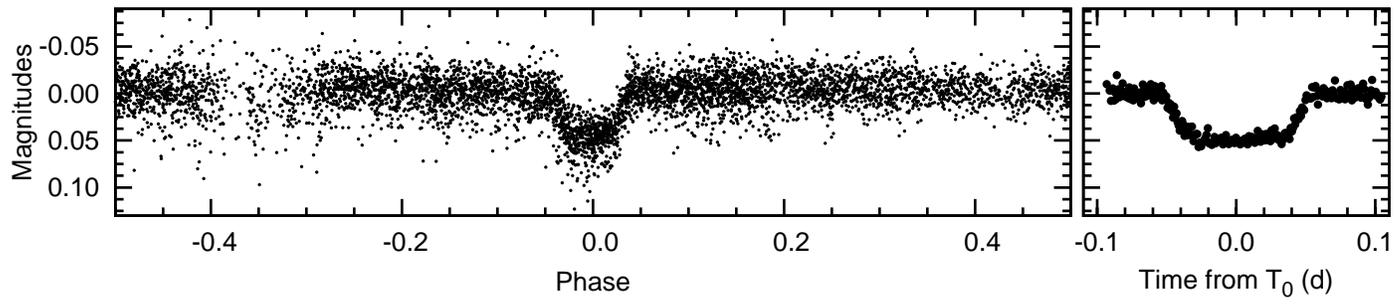

# Towards the Rosetta Stone of planet formation

G.Maciejewski[1], R.Neuhäuser[1], R.Errmann[1], M.Mugrauer[1],
Ch.Adam[1], A.Berndt[1], T.Eisenbeiss[1], S.Fiedler[1], Ch.Ginski[1],
M.Hohle[1,2], U.Kramm[3], C.Marka[1], M.Moualla[1], T.Pribulla[1],
St.Raetz[1], T.Roell[1], T.O.B.Schmidt[1], M.Seeliger[1],
I.Spaleniak[1], N.Tetzlaff[1] & L.Trepl[1]

[1] *Astrophysikalisches Institut und Universitäts-Sternwarte,
Schillergässchen 2-3, D-07745 Jena, Germany*
[gm@astro.uni-jena.de]

[2] *Max-Planck-Institut für Extraterrestrische Physik,
Giessenbachstraße, 85741 Garching, Germany*

[3] *Institut für Physik, Univ. Rostock, D-18051 Rostock, Germany*

**Abstract.** Transiting exoplanets (TEPs) observed just ∼10 Myrs after formation of their host systems may serve as the Rosetta Stone for planet formation theories. They would give strong constraints on several aspects of planet formation, e.g. time-scales (planet formation would then be possible within 10 Myrs), the radius of the planet could indicate whether planets form by gravitational collapse (being larger when young) or accretion growth (being smaller when young). We present a survey, the main goal of which is to find and then characterise TEPs in very young open clusters.

## 1. Introduction

Detection of flat-bottomed eclipses with the depth of a few tens of millimagnitudes in a light curve of a star suggests the presence of a transiting sub-stellar mass object. The depth of a transit together with the stellar radius give some constraints on the radius of the transiting object. However, no information on the mass of this companion can be extracted from photometric measurements. It may be a planet, a brown dwarf or a low-mass star, because in the low-mass regime the radius is independent of mass (Guillot 1999). A Doppler follow-up is the only way to determine mass and, hence, true nature of the transiting object. Spectroscopic observations of a central star, which are a by-product of the radial velocity measurements, are necessary to better determine, inter alia, the radius of the star and, in turn, that of the companion. Mass and radius lead to a measurement of the transiting object mean density - an essential parameter for the study of the internal structure of extrasolar planets, brown dwarfs and low-mass stars.

## 2. Trumpler 37 campaign

We search for variable stars and transiting exoplanet candidates among young clusters by monitoring them photometrically for several weeks in all clear hours and nights using the CCD-imager STK (Mugrauer & Berthold 2010) at the 90cm telescope of the University Observatory Jena. In 2009 we initiated our survey by observing the Trumpler 37 cluster, the age of which is estimated to be 1-5 Myrs (Sicilia-Aguilar *et al.* 2004). We obtained light curves for 18,000 stars in Trumper-37. We found dozens of variable stars. For 35 of those stars, we see light curves typical for eclipsing binaries. We also measured rotation periods for many cluster members and detected several flares among members. Since 2010 we have organised an international campaign to observe for 24 hours each day and night for a certain number of days/nights not to miss any transiting planets due to limited observing time. Telescopes located at different longitudes have been engaged (to be reported elsewhere).

In the 2009 campaign, we found that the light curve of a 15-mag star reveals features typical for a planetary transit. Every 1.36 day its brightness drops by $47.2 \pm 1.2$ mmag causing a flat-bottom flux dip (Fig. 1). The secondary eclipse is not detectable. Based on *BVJHK* photometry, we determined the host star's spectral type to be G8. We noted that the star is located close to the main sequence of the Trumpler 37 cluster. The proper motion of the star is also consistent with cluster membership, so it may be preliminary treated as a cluster member.

The transit depth indicates a radius of the transiting object of about 2 $R_{\rm J}$, and hence mass of about 15 $M_{\rm J}$ at 5 Myrs, following evolutionary models of cool brown dwarfs and extrasolar giant planets formation (e.g. Baraffe *et al.* 2003). A variety of phenomena may mimic transit light curves. Examples of these false positives are the central transit of a low-mass star in front of a large main-sequence star or red giant, grazing eclipses in systems consisting of two main-sequence stars and a contamination of a fainter eclipsing binary along the same line of sight. Using photometric and spectroscopic methods we try to rule out these false-positive scenarios in the near future.

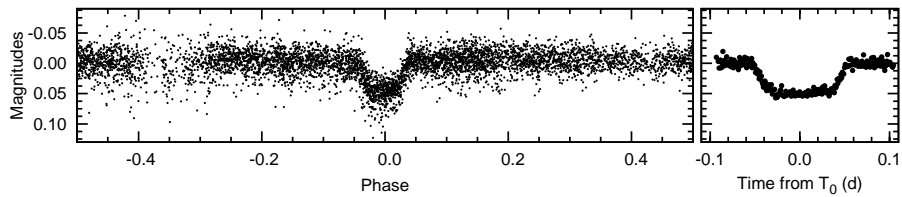

Figure 1.: *Left: The phased light curve of the planetary transit candidate composed of 5548 individual measurements collected in the R-band filter with the 90-cm telescope near Jena, Germany. Right: First results of the photometric follow-up: the transit-candidate I-band light curve obtained with the 2.2-m telescope at Calar Alto (Spain) in July 2010.*

Our exoplanet candidate would be by far the youngest transiting planet known and the youngest known exoplanet candidate at all. If confirmed to be a planet, it will give strong constraints on several aspects of planet formation. The planet's radius could indicate whether this planet formed by gravitational collapse or accretion growth. It would be the youngest transiting exoplanet with several highly important parameters known (such as mass, radius, density, age, and distance), which could serve as the Rosetta Stone for planet formation theories.

*Acknowledgements.* GM and TP acknowledge support from the EU in the FP6 MC ToK project MTKD-CT-2006-042514. GM and RN acknowledge support from the DAAD PPP–MNiSW project 50724260–2010/20011. RN would like to acknowledge general support from DFG in programmes NE 515/23-1, 30-1 and 32-1. AB and SR thank the DFG for support in the program NE 515/32-1. CM, CG, UK, TS and LT thank the DFG for support in programs SCHR 665/7-1, NE 515/30-1, SPP 1385, NE 515/30-1 and SFB TR 7, respectively. M.Moualla thanks the Syrian government for their support. NT acknowledges financial support from Carl-Zeiss-Stiftung.